\documentclass{agnspec}
\usepackage{psfig}

\def\pn{\par\noindent}
\def\chandra{{\it Chandra}}
\def\pn{\par\noindent}
\def\xmm{XMM--{\it Newton}}
\def\cgs{erg cm$^{-2}$ s$^{-1}$} 
\begin{document}
\title{An XMM--{\it Newton} survey of Extremely Red Objects}
\author{M. Brusa\inst{1}, A. Comastri\inst{2}, E. Daddi\inst{3},
A. Cimatti\inst{4}, C. Vignali\inst{5}}
\institute{Dipartimento di Astronomia, Universit\`a di Bologna,
via Ranzani 1, I-40127 Bologna, Italy
\and INAF --  Osservatorio Astronomico di Bologna, via Ranzani 1, 
I--40127 Bologna, Italy
\and European Southern Observatory, Karl--Schwarzschild--Strasse 2,
D--85748 Garching bei Muenchen, Germany
\and INAF -- Osservatorio Astrofisico di Arcetri, Largo E. Fermi 5, 
I--55025 Firenze, Italy
\and Dept. of Astronomy and Astrophysics, 
 The Pennsylvania State University, 525 Davey Lab, 
University Park, PA 16802, USA}

\authorrunning{Brusa M. et al.}

\maketitle

\begin{abstract}
We present preliminary results of a 25 ks \xmm\ observation of the 
largest sample of 
Extremely Red Objects (EROs) available to date ($\sim$450 sources), 
selected in a contiguous area of $\sim 700$ arcmin$^2$ (Daddi et al. 2000). 
Five of the 36 hard X--ray selected sources brighter than 
$7\times 10^{-15}$ \cgs\ in
the 2--10 keV band,  are associated
with EROs.
All of the X--ray detected EROs show rather extreme X--ray--to--optical
flux ratios, suggesting the presence of highly obscured AGN activity.

\end{abstract}

\section{Introduction}

{\bf E}xtremely {\bf R}ed {\bf O}bjects (EROs, $R-K>5$) were
discovered serendipitously a decade ago (Elston et al. 1988)
from optical and near--infrared selection criteria.  Having the
colors expected for high-redshift passive ellipticals, EROs can be used as
tracers of distant and old spheroids as a test for different cosmological models.\\
Recent wide-area surveys have shown that the surface density of EROs is
consistent with that expected for elliptical galaxies, assuming pure
luminosity evolution (PLE). About 10\%
of all objects down to a $K$-band magnitude limit of 19.2 are 
EROs (Daddi et al. 2000; Daddi, Cimatti 
\& Renzini, 2000). Furthermore, a large angular clustering signal
associated with EROs 
was detected from the $K$--selected sample over the same area,
that is due to 
the large--scale structure traced by early--type galaxies
at z$\sim1$ (Daddi et al. 2002).\\    
On the other hand, the broad--band properties of EROs are also consistent with 
those expected for high-redshift dusty starbursts 
(e.g.,\ HR--10, Cimatti et al. 1998; Smail et al. 1999) and Active 
Galactic Nuclei (AGN) reddened by dust (Pierre et al. 2001).\\
The relative fraction of old ellipticals, 
dusty star--forming systems and obscured AGN among optically selected 
ERO samples is a key parameter in the study of the galaxy evolution
and the link between massive elliptical galaxy in formation and 
the onset of AGN activity (Almaini et al. 2002; Granato et al. 2001).\\
\pn
A step towards constraining the relative fractions of these different
source types within the ERO population has been recently achieved by
the {\tt $K20$ survey} (Cimatti et al. 2002; Daddi et
al. 2002) by means of deep VLT spectroscopy. 
The spectroscopic counterparts of about 45
EROs with $K\leq 19.2$ turned out to be in about equal proportion 
old, passive ellipticals (31\%) and dusty emission--line galaxies
(33\%) at $z=0.8-1.5$, while a comparable
fraction is still unidentified.  
Interestingly, the average spectrum of the emission--line EROs 
shows a very red and smooth continuum with
[OII]$\lambda$3727 and a weak [NeV]$\lambda$3426, suggesting
that a fraction of these sources hosts hidden AGN activity. \\
Given that the Spectral Energy Distribution
(SED) of optically obscured type II AGN 
is dominated by the host galaxy starlight, they show $R-K$ 
colors similar to those of passive ellipticals; therefore,  
sensitive hard X--ray observations provide a powerful tool 
to uncover AGN activity within the ERO population.
Indeed, a sizeable fraction of the new population of faint hard X--ray 
selected sources, recently investigated in \chandra\ and \xmm\ 
medium--deep observations, turned out to be faint 
($R>24.5$), obscured objects, with red optical to near--infrared 
colours (Mushotzky et al. 2000; 
Cowie et al. 2001; Gandhi et al. 2001; Hornschemeier et al. 2001; 
Alexander et al. 2002; Mainieri et al. 2002a). \\
\pn
The most extensive study on the X--ray properties of EROs to date was
performed by Alexander et al. (2002) using the 1 Ms \chandra\ Deep Field
North (CDF-N) observation. Targeting a 70.3 arcmin$^2$ region close
to the CDF-N aim-point, Alexander et al. (2002) detected X--ray emission
from 13 EROs ($21^{+12}_{-8}$\% of the \hbox{$K\le20.1$} ERO sample). 
The majority (2/3) 
of the X--ray detected EROs were found to have X--ray
properties consistent with highly obscured AGNs [\hbox{$N_{\rm
H}\approx$~(5--35)$\times10^{22}$ cm$^{-2}$} for $z=$~1--3]. 
Moreover, starbursts and normal elliptical galaxies were also 
possibly detected at the faintest X--ray fluxes. 
Although having deep X--ray and optical
observations, this study is limited in areal coverage, and 
therefore is unsuitable for detailed statistical and clustering 
analyses of the ERO population.\\


\pn
We have started an extensive program of multiwavelength 
observations of the largest sample of EROs available to date
($\sim$450 sources), selected in a contiguous area over a $\sim 700$
arcmin$^2$ field (the ``Daddi field'', Daddi et al. 2000) and
complete to a magnitude limit of K$=$19.2. 
Deep optical (R$\sim 26.2$ at the 3$\sigma$ level) photometry is 
available for the present field 
(see Daddi et al. 2000 for details in optical and 
near--infrared data reduction) and VIMOS spectroscopy is 
planned at VLT.
Here we present preliminary results from the first 25 ks \xmm\ observation of 
this field (Sec. 2) and compare our results with other recent findings
from medium and deep hard X--ray surveys (Sec. 3). Finally, we try to
draw some conclusions on the nature of hard X--ray selected EROs
(Sec. 4).
\begin{figure}
\centerline{\psfig{figure=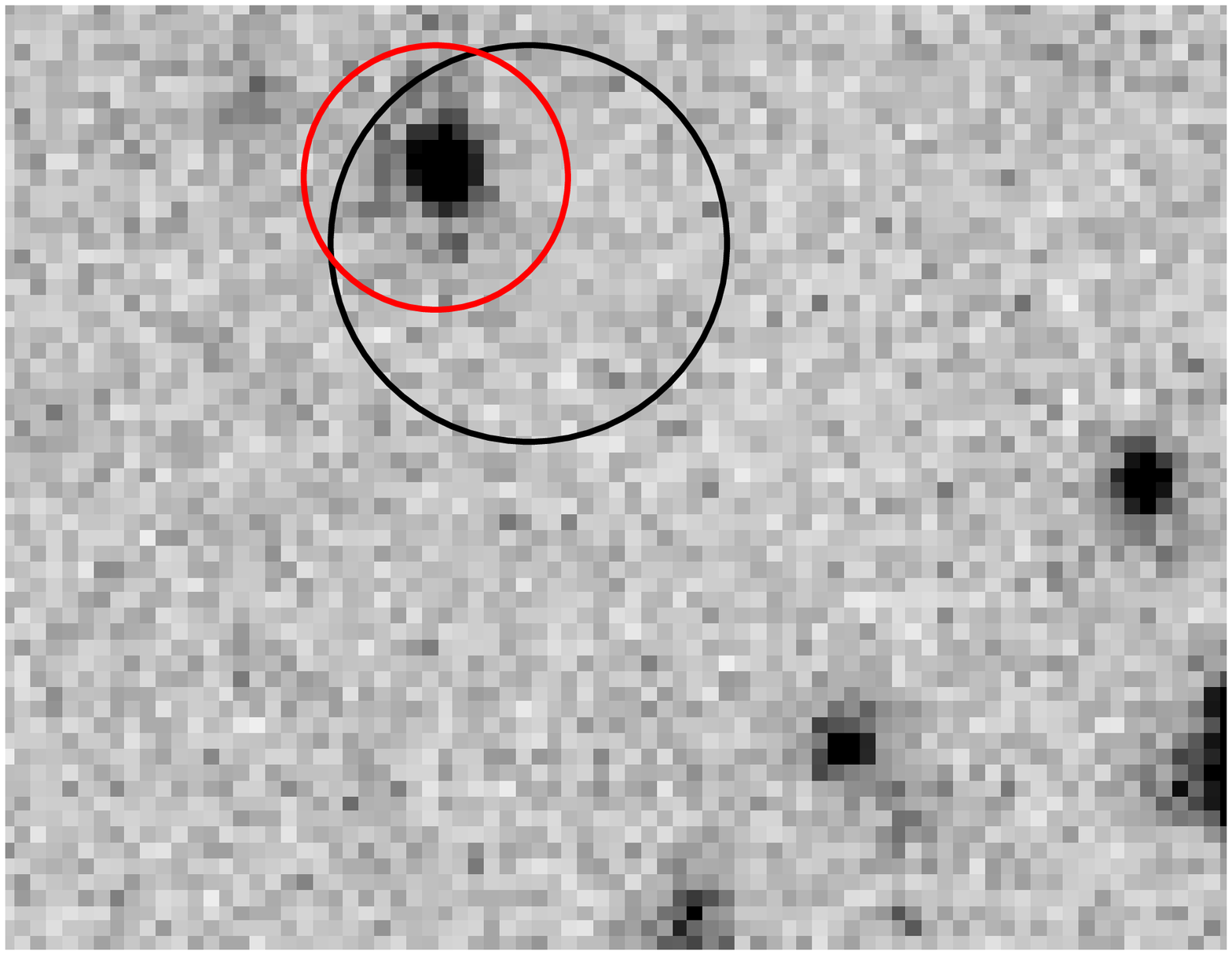,width=5.5cm,angle=0}}
\centerline{\psfig{figure=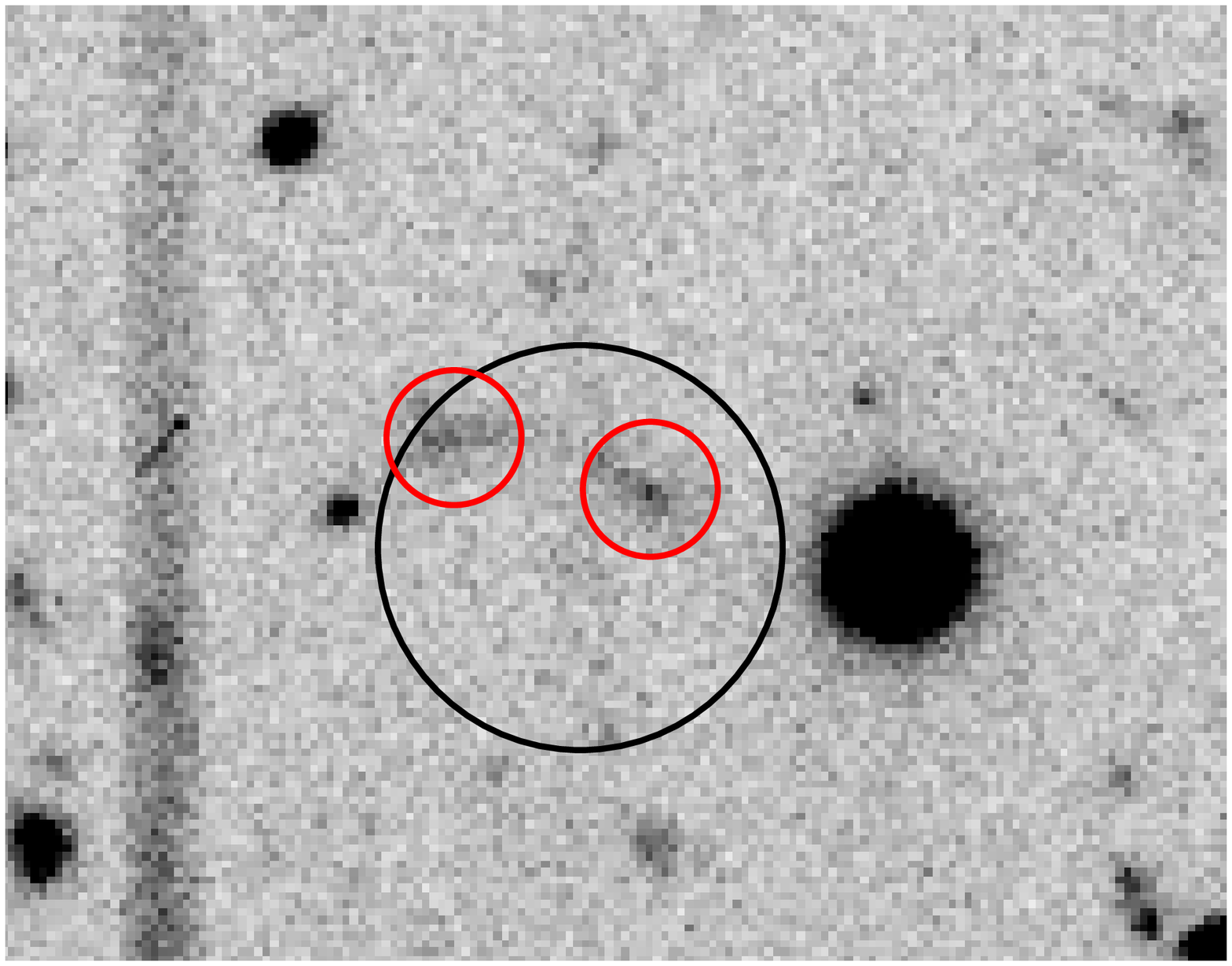,width=5.5cm,angle=0}}
\caption[]{R--band optical image with superimposed 
the \xmm\ error circle. {\it Upper panel}: a secure identification
of an X--ray source on--axis: the \xmm\ error circle radius (large circle) is 
3$''$, while the ERO counterpart is indicated by the small circle (2$''$). 
{\it Lower panel}: the ambiguous identification of an X--ray source
off--axis: the \xmm\ error circle has a 6$''$ radius, while 
the positions of two possible counterparts (both of them EROs) 
are indicated with the smaller 2$''$ circles.} 
\end{figure}

\section{Data reduction and results}

The Daddi field was observed on August 3, 2001 for a nominal exposure
time of $\sim 50$ ks.\\
The \xmm\ data were processed using version 5.2 of the Science Analysis System 
(SAS). The event files were cleaned up from hot pixels and soft proton
flares, removing all the time intervals with a count rate higher than 0.15
c/s in the 10--12.4 keV energy range for the two MOS and higher than 0.35
c/s in the 10--13 keV band for the {\it pn} unit.
The resulting exposure times are $\sim 26$ ks in the MOS1 and MOS2
detectors and $\sim 30$ ks in the {\it pn} detector.\\
The {\em EBOXDETECT} task, the standard {\em SAS} sliding box cell
detect algorithm, was run on the 2--10 keV cleaned event. 
Thirty--six sources were detected above the 3.5$\sigma$
level, corresponding to a flux 
limit of $\sim 7 \times 10^{-15}$ \cgs assuming a power--law spectrum
with $\Gamma$=1.8 and Galactic absorption (N$_H=5\times10^{20}$ cm$^{-2}$).\\ 
The X--ray centroids have been corrected for systematic errors with
respect to the optical positions of three bright quasars already known
in the field (from NED database) and then were cross correlated with 
the $K$--band catalog.
\begin{figure}
\centerline{\psfig{figure=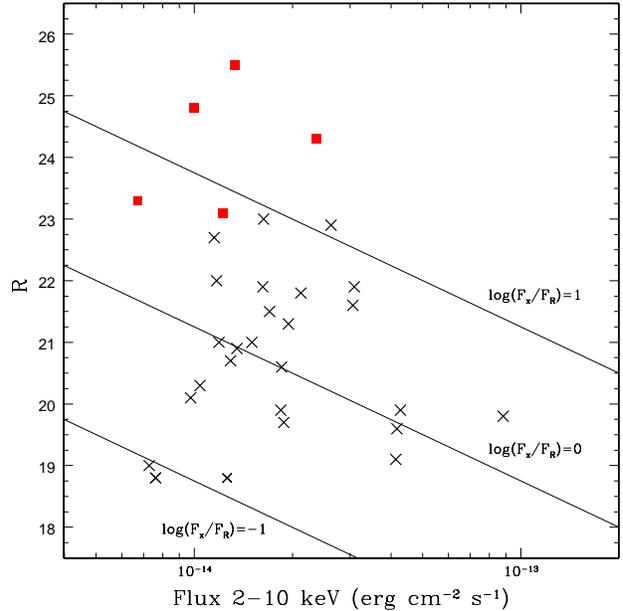,width=8.8cm,angle=0}}
\caption[]{R--band magnitude vs. hard X--ray flux for the 31 sources
with $K$--band counterparts detected
in the XMM--{\it Newton} observation. Filled squares are EROs. The
solid lines represent loci of equal X--ray--to--optical flux ratio
(0.1, 1, 10).}
\end{figure}
\pn
We searched for 
near--infrared counterparts within a radius of
3$''$ from the X--ray position in the inner region of the \xmm\
field--of--view (10 arcmin), while for the objects detected 
at larger off--axis angles we adopted a searching radius of $6''$ 
to take into account the broadening  of the \xmm\ PSF at increasing 
distance from the aim point. 
All but five of the hard X--ray selected 
sources have at least one counterpart in the K--band image.
The number of expected spurious sources, computed from 
K--band galaxy surface density (Daddi et al. 2000) is about 3. 
We found a unique candidate for 24 sources 
(4 of them with R--K $>$ 5), while for the remaining 7 objects 
(most of them at large off--axis angles)
we detected two possible counterparts. 
In one of these cases the 2 optical counterparts 
have similar R band magnitudes and EROs colors (Fig.~1); 
for the purposes of the present paper
we consider the nearest ERO as the counterpart of the X--ray
source.\\
Five of the 31 hard X--ray selected sources with near--infrared counterparts 
are thus tentatively associated with EROs.
A more detailed discussion on the \xmm\ data analysis, the 
associations of X--ray sources with optical counterparts and 
the analysis of X--ray colors is reported in Brusa et al. (2002).\\ 
\pn
The hard X--ray and optical fluxes of the X--ray sources 
with a $K$--band counterpart are reported in Fig. 2. 
All the EROs detected in the \xmm\ observation show relatively high
X--ray--to--optical flux ratios (f$_{\rm X}$/f$_{\rm O}$$\sim$10)
compared to those of the other X--ray detected sources.
\pn
The \xmm\ field of view covers an area including $\sim350$ EROs: 
the fraction of X--ray active EROs in this optically selected 
sample is therefore $\sim1.5$\% (5/350);
on the other hand, the fraction of hard X--ray sources with EROs colors 
is much higher, $\sim 15\%$ (5/36), and is in agreement with other 
\xmm\ findings (Lehmann et al. 2001).

\section{Comparison with literature samples}

In order to investigate the nature of hard X--ray selected EROs
and the link between faint hard X--ray sources and the 
ERO population, we have collected all the multiwavelength data
available in literature for EROs serendipitously discovered both in 
\xmm\ and \chandra\ medium--deep observations:
\begin{itemize}
\item[$\bullet$] Thirteen EROs 
detected in the 2--8 keV band in the \chandra\ Deep Field North observation 
(Alexander et al. 2002\footnote{These 
authors adopted different criteria in the EROs definition: $I-K>4$,
which roughly correspond to $R-K>5.3$.}, Hornschemeier et al. 2001);
\item[$\bullet$] Five hard X-ray selected EROs
serendipitously detected in medium--deep \chandra\ observations 
in the fields of SSA13 (Mushotzky et al. 2000), A370 (Barger et al. 2001), and 
A2390 (Cowie et al. 2001, Crawford et al. 2001);
\item[$\bullet$] Eleven EROs detected in the hard (2--10 keV) band in
the \xmm\ Lockman Hole observation (Mainieri et al. 2002b)
\end{itemize}
\pn
The total sample consists of 34 objects, including the 5 EROs
discussed in the present work.\\
The R--band magnitudes plotted versus the 2--8 keV fluxes are reported 
in Fig.~3 .
The X--ray fluxes have been converted into the 
2--8 keV energy range, assuming the spectral models quoted
in the literature, while the R--band magnitudes 
have been obtained from published I--band magnitudes 
adopting $R-I$=1.3 (Alexander et al. 2002).\\
\pn
\begin{figure}
\centerline{\psfig{figure=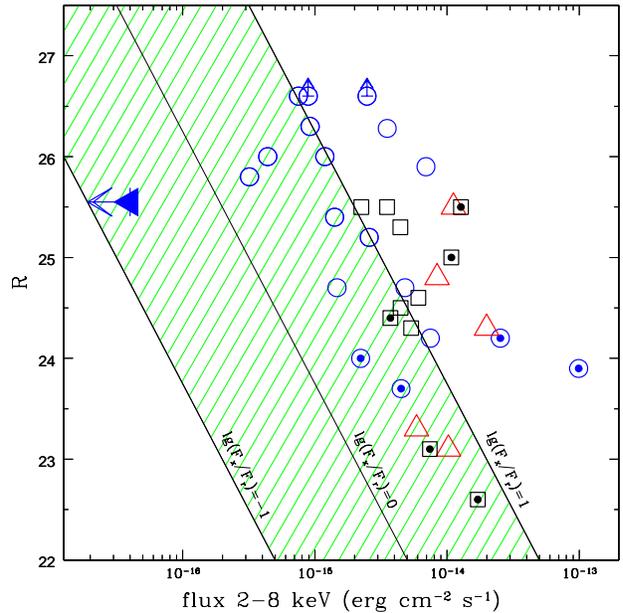,width=8.8cm,angle=0}}
\caption[]{R--band  magnitude vs. hard X--ray flux for a sample of literature
EROs, serendipitously detected in hard X--ray surveys. Circles
refer to \chandra\ observations, squares to \xmm\ Lockman Hole
observation, and triangles to the present data. 
Dot--filled symbols are identified objects (mostly highly obscured,  
high--luminosity Type II AGN). Open symbols correspond to
unidentified objects. For comparison, we report the result of stacking 
analysis performed on all the EROs in the 
HDF--N field not individually detected in the \chandra\ observation 
(Alexander et al. 2002; upper limit at the faintest X--ray flux).}
\end{figure}
The present EROs sample span a wide range of optical and
X--ray fluxes: all of the objects show a relatively well-defined 
correlation between the X--ray fluxes and the optical magnitude 
around f$_{\rm X}$/f$_{\rm O}\simeq$ 10. This correlation is 
shifted by one order 
of magnitude from the one found by {\tt ROSAT} for soft X--ray 
selected quasars (Hasinger et al. 1998) and recently extended 
by {\it Chandra} and XMM--{\it Newton} observations also for 
hard X--ray selected sources (Alexander et al. 2002; Lehmann et al.
2001).\\
\pn
The most plausible explanation of such a high X--ray--to--optical flux
ratio in these objects is the presence of 
high obscuration towards the 
active nuclear source. \\
This hypothesis is strongly supported both by X--ray spectral analysis
(Gandhi et al. 2001, 2002; Alexander et al. 2002; Cowie et al. 2001;
Mainieri et al. 2002a,b) and by the optical identifications available for 
9 objects of this sample: all but one of the sources 
are identified with Type II AGN at $z>0.7$, either from optical 
spectroscopy (4 objects) or from deep multi--band photometry (4 objects) 
(Cowie et al. 2001; Mainieri et al. 2002b; Hornschemeier et al. 2001). 
The only source not classified as a type II AGN is the z=1.020 
emission--line galaxy in the Hornschemeier et al. sample 
(see their Table 10).

\section{Discussion and conclusions}

All the findings discussed above support the idea that 
a significant fraction of the optical counterparts of hard X--ray
selected sources are EROs.
Indeed, the fraction of EROs among the hard X--ray sources in the
Daddi field is $\sim$15\%; this value rises up to $\sim 25\%$ 
in the deeper (F$_{2-10 keV}\sim 2\times 10^{-15}$ \cgs) \xmm\ survey 
of the Lockman Hole, and has to be
considered as a lower limit, as several X--ray sources are not covered
by deep near--infrared observations. 
Interestingly enough, if we consider only the sources 
with f$_{\rm X}$/f$_{\rm O}>10$, the fraction of EROs 
is even higher 
(more than 50\% in the Lockman Hole observation, Mainieri et al. 2002a). 
The so far identified EROs have high X--ray luminosity 
(L$_X$$>$10$^{44}$ erg s$^{-1}$) and they actually are heavily obscured AGN,
as inferred from X--ray spectral analysis  
(Cowie et al. 2001; Gandhi et al. 2002; Mainieri et al. 2002a). 
Therefore, hard X--ray selected EROs (or at least a fraction of them) 
have properties similar to those of Quasars 2, the high--luminosity, 
high redshift type II AGNs predicted in X--Ray Background (XRB) synthesis
models (e.g. Comastri et al. 1995). \\
Deep optical and near--infrared follow--up of complete samples of hard X--ray
selected sources with extreme f$_{\rm X}$/f$_{\rm O}$ will 
definitively assess the fraction of reddened sources among the 
XRB constituents. \\
\pn 
On the other hand, the fraction of hard X--ray sources within optically
selected ERO samples is much lower. At the optical and
X--ray flux limits of our survey, the fraction is $\sim$1.5\%
and rises up to $\sim15$\% in the deep 1 Ms CDF--N survey 
(Alexander et al. 2002). However, it is unlikely that this fraction 
is much higher than this value. Indeed, the stacking analysis of 
EROs in the HDF--N actually confirms the lack of hard X--ray emission 
at fluxes $<4\times10^{-17}$ \cgs, with an average f$_{\rm X}$/f$_{\rm
O}$ value which is at least two order of magnitudes lower than the average
ratio of the individually X--ray detected EROs (Fig.~3).
The upper limit on the average f$_{\rm X}$/f$_{\rm O}$ value is 
consistent with that of passive evolving ellipticals or dusty 
starburts suggesting the lack of strong AGN activity in the majority
of optically selected EROs.\\
Deeper X--ray observations of large samples
of $K$--selected EROs would be crucial to compute the fraction of 
X--ray active EROs on the widest area possible (to avoid cosmic
variance). A detailed study of the clustering properties of these objects
would shed new light on the link between nuclear activity and galaxy 
evolution.

\begin{acknowledgements}
This research has been partially supported by ASI contracts
I/R/103/00, I/R/107/00 and I/R/27/00, and by the MURST grant 
Cofin-00--02--36.
CV also thanks the NASA LTSA grant NAG5--8107 for financial support.
We gratefully thank D. Alexander, G. Zamorani and F. Fiore for useful 
discussions and V. Mainieri for providing data prior to publication.
\end{acknowledgements}

\end{document}